\documentclass[matzei,envcountsame,envcountsect]{svjour}

\usepackage{amsmath,amssymb}

  \usepackage{for-arXiv}

\makeatletter

\providecommand{\one}{\leavevmode\hbox{\upshape{\small1\kern-3.35pt\normalsize1}}}%
\def\l2zd{\ell^{2}(\mathbb{Z}^{d})}
\def\zd{\mathbb{Z}^{d}}
\def\DN{\Delta_{\mathrm{N}}}
\def\DD{\Delta_{\mathrm{D}}}
\def\DX{\Delta_{\mathrm{X}}}
\def\DDT{\Delta_{\,\widetilde{\mathrm{D}}}}
\def\XinXset{\mathrm{X}\in\{\mathrm{N},\widetilde{\mathrm{D}},\mathrm{D}\}}
\def\NX{N_{\mathrm{X}}}
\def\NN{N_{\mathrm{N}}}
\def\ND{N_{\mathrm{D}}}
\def\NDT{N_{\widetilde{\mathrm{D}}}}
\def\PX{P_{\mathrm{X}}}
\def\PN{P_{\mathrm{N}}}

\def\EX{E^{(1)}_{\mathrm{X}}}
\def\EN{E^{(1)}_{\mathrm{N}}}
\def\ED{E^{(1)}_{\mathrm{D}}}
\def\EDT{E^{(1)}_{\widetilde{\mathrm{D}}}}

\def\e{\mathrm{e}}
\def\d{\mathrm{d}\hspace{1pt}}
\def\Chi{\raisebox{.4ex}{$\chi$}}

\def\limsup{\mathop\mathrm{lim\,sup}}

\def\spec{\mathop\mathrm{spec}}
\def\tr{\mathop\mathrm{trace}}
\def\ooplus{\mathop{\raisebox{-.28ex}{\Large$\oplus$}}}

\def\le{\leqslant}
\def\ge{\geqslant}

\def\@biblabel#1{[#1]}

\newif\ifper\pertrue

\def\au#1#2{#2, #1}
\def\et{,\ }
\def\ti#1{: #1.}

\def\bti{\@ifnextchar[\bbti\bbbti}
\def\bbti[#1]#2{: #2, #1}
\def\bbbti#1{: #1.}

\def\z{\@ifnextchar[\zz\zzz}
\def\zz[#1]#2#3#4#5{\perfalse{#2} \textbf{#3}, #4 (#5) [#1]}
\def\zzz#1#2#3#4{{#1} \textbf{#2}, #3 (#4)\ifper\fi\pertrue}

\def\pub{\@ifstar\pubstar\pubnostar}
\def\pubnostar{\@ifnextchar[\@@pubnostar\@pubnostar}
\def\@@pubnostar[#1]#2#3#4{#2, #3, #4, #1\ifper\fi\pertrue}
\def\@@pubnostar[#1]#2#3#4{#2, #3, #4, #1}
\def\@pubnostar#1#2#3{#1, #2, #3\ifper\fi\pertrue}
\def\pubstar[#1]#2#3#4{\perfalse #2, #3, #4 [#1]}

\newcounter{numcount}
\newcommand{\labelnummer}{\mbox{\normalfont (\roman{numcount})}}%

\newenvironment{nummer}%
  {\let\curlabelspeicher\@currentlabel%
    \begin{list}{\labelnummer}{\usecounter{numcount}\leftmargin0pt%
        \topsep0.5ex\partopsep2ex\parsep0pt\itemsep1ex\@plus1\p@%
        \labelwidth3.5em\itemindent4.5em\labelsep1em}%
      \let\saveitem\item%
      \def\item{\saveitem%
        \def\@currentlabel{\curlabelspeicher\labelnummer}%
        \let\label\bemlabel}}%
    {\end{list}}%
  
  \def\itemref#1{\expandafter\@setref\csname r@#1item\endcsname%
                  \@firstoftwo{#1}}%
  
  \def\bemlabel#1{\@bsphack%
    \protected@write\@auxout{}%
    {\string\newlabel{#1}{{\@currentlabel}{\thepage}}}%
    \ifmmode\else%
    \protected@write\@auxout{}%
    {\string\newlabel{#1item}{{\labelnummer}{\thepage}}}%
    \fi%
    \@esphack}%

\spnewtheorem{remarks}[theorem]{Remarks}{\itshape}{\rmfamily}

\@addtoreset{equation}{section}

\let\saveref\ref
\def\ref#1{\textup{\saveref{#1}}}

\makeatother

\begin{document}

\title{Spectral properties of the Laplacian on bond-percolation graphs} 

\author{Werner Kirsch$\,$\thanks{\email{werner.kirsch@rub.de}} 
  \and
  Peter M\"uller$\,$\thanks{\emph{On leave from:} Institut f\"ur Theoretische
    Physik, Georg-August-Universit\"at, Friedrich-Hund-Platz~1, 
    \mbox{D--37077 G\"ottingen}, Germany;\\
    \email{peter.mueller@\discretionary{physik.}{}{physik.}uni-goettingen.de}}
  }


\institute{Fakult\"at und Institut f\"ur Mathematik, 
  Ruhr-Universit\"at Bochum, Universit\"ats\-stra{\ss}e 150,
  \mbox{D--44780 Bochum}, Germany
  } 

\def\today{12 December 2005}
\date{Version of \today}

\maketitle

\begin{abstract}
  Bond-percolation graphs are random subgraphs of the $d$-dimensional
  integer lattice generated by a standard bond-percolation process.  The
  associated graph Laplacians, subject to Dirichlet or Neumann conditions at
  cluster boundaries, represent bounded, self-adjoint, ergodic random
  operators with off-diagonal disorder. They possess almost surely the
  non-random spectrum $[0,4d]$ and a self-averaging integrated density
  of states. The integrated density of states is shown to exhibit Lifshits
  tails at both spectral edges in the non-percolating phase. While the
  characteristic exponent of the Lifshits tail for the Dirichlet (Neumann)
  Laplacian at the lower (upper) spectral edge equals $d/2$, and thus depends
  on the spatial dimension, this is not the case at the upper (lower) spectral
  edge, where the exponent equals $1/2$.

\end{abstract}


\vspace{.2cm}

%
\section*{Introduction}
%

Spectral graph theory studies linear operators which are associated
with graphs.  The goal is to see how properties of the graph are
reflected in properties of the operators and vice versa. This has
attracted vivid interest in the last two decades
\cite{Moh91,CvDo95,Chu97,Col98}. The kind of graphs we shall be
concerned with in this paper are bond-percolation graphs \cite{Gri99},
a special type of random subgraphs of the $d$-dimensional integer
lattice. They are of popular use in Physics for modelling various
types of random environments \cite{StAh94,BuHa96}. On these graphs we
consider Laplacians with different kinds of boundary conditions at
cluster borders and study some of their spectral properties. Apart
from the non-randomness of the spectrum, and of the spectral
components in the Lebesgue decomposition, we show the existence and
self-averaging of the integrated density of states. The main result
establishes Lifshits tails of the integrated density of states at the
lower and upper spectral edge in the non-percolating phase. Depending
on the boundary condition and on the spectral edge, the Lifshits tail
discriminates between stretched (i.e.\ linear) and condensed (i.e.\
cube- or ball-like) clusters which contribute the dominating
eigenvalues. The crucial technical estimates in our proof are Cheeger
\cite{Col98} and Faber-Krahn \cite{ChGr00} isoperimetric inequalities
on graphs. Our analysis here is facilitated by the fact that in the
non-percolating phase almost all graphs consist of infinitely many
finite clusters. Yet, the non-percolating phase gives rise to
interesting phenomena, because it supplies clusters of arbitrarily
large size. In the percolating phase one has to cope with the infinite
cluster, too, which requires a more intricate understanding.  This
case will be studied in \cite{MuSt05}. Related previous work can be
found in \cite{Ant95,BiKo01}.

Spectral properties of Laplacians on bond-percolation (or related) graphs have
been studied in the Physics literature, see the general accounts
\cite{StAh94,BuHa96} or the recent examples \cite{BiMo99,BrLo01,Mul03} for
applications to soft matter. The Lifshits tails, whose existence we prove
here, were sought after in the numerical simulations \cite{BrAs01} for the
Neumann Laplacian on two- and three-dimensional bond-percolation graphs. The
tails could not be observed there due to finite-size corrections and the
considerable numerical effort needed to access such rare events. For the
different case of Erd\H{o}s--R\'enyi random graphs, however, the existence of
Lifshits tails for the Neumann Laplacian was known on the basis of analytical,
non-rigorous arguments \cite{BrRo88,BrAs01}, which inspired our proof here.
Moments of the eigenvalue density for this model were rigorously analysed in
\cite{KhSh04}. Other models in the physics literature deal with the adjacency
operator on bond-percolation graphs \cite{KiEg72,ShAh82}. Quite often, this
goes under the name quantum percolation. Yet, from a rigorous mathematical
point of view, all of the above models with off-diagonal disorder have
remained widely unexplored, see however \cite{Ant95}.

In contrast, Laplacians on \emph{site}-percolation graphs of the
$d$-dimensional integer lattice belong to the class of models with
diagonal disorder. Therefore they are closer to the range of
applicability of the highly developed theory of random Schr\"odinger
operators \cite{Kir89,CaLa90,PaFi92,Sto01,LeMu03}. This is partly of
help for analysing their spectral properties with mathematical rigour.
For finite-range hopping operators on site-percolation graphs, the
non-randomness of the spectrum and existence of the integrated density
of states was shown in \cite{Ves05a,Ves05b}.  Particular emphasis is
laid on the behaviour of the spectrum related to finitely supported
eigenfunctions, see also \cite{ChCh86}, where the issue was first
taken up from a mathematical point of view.  Furthermore, a Wegner
estimate is established in \cite{Ves05b} for an Anderson model on
site-percolation graphs. Highly developed large-deviation techniques
for the parabolic Anderson model are used in \cite{BiKo01} to prove
Lifshits tails for the integrated density of states of the Laplacian
$\DDT$ (in the sense of our Definition in Eq.\ \eqref{DTdef} below) on
\emph{site}-percolation graphs.

This paper is organised as follows. In Section~\ref{defres} we give the
precise definitions of the objects we are dealing with and state our results.
Theorem~\ref{main} in Subsection~\ref{results} contains the central result on
Lifshits tails of the integrated density of states. All proofs are deferred to
Section~\ref{proofs}.

%
\section{Definitions and Results}  \label{defres}
%

%
\subsection{Bond-percolation graphs}
%

A thorough and comprehensive account of (bond) percolation can be found in
Grimmett's textbook \cite{Gri99}, which serves as a standard reference on the
subject. For $d \in\mathbb{N\,}$, a natural number, we denote by
$\mathbb{L}^{d}$ the (simple hypercubic) lattice in $d$ dimensions. Being a
graph, the lattice $\mathbb{L}^{d} = (\zd, \mathbb{E}^{d})$ has a \emph{vertex
  set}, 
which consists of the $d$-dimensional integer numbers $\mathbb{Z}^{d}$, and an
\emph{edge set} $\mathbb{E}^{d}$ given by all unordered pairs $[x,y]$ of
nearest-neighbour vertices $x,y \in \zd$, that is, those vertices which have
Euclidean distance $|x-y| : = \bigl(\sum_{\nu=1}^{d} |x_{\nu}
-y_{\nu}|^{2}\bigr)^{1/2} = 1$. Here, elements of $\zd$ are canonically
represented as $d$-tuples $x=(x_{1},\ldots, x_{d})$ with entries from
$\mathbb{Z}$.

Given any subset of vertices $\varnothing \neq \mathcal{V} \subseteq
\mathbb{Z}^{d}$ and a subset of edges $\mathcal{E}\subseteq \{ [x,y]
\in \mathbb{E}^{d} : x,y \in\mathcal{V} \}$ between them, we call the
graph $\mathcal{G} := (\mathcal{V}, \mathcal{E})$ a \emph{subgraph} of
$\mathbb{L}^{d}$. The \emph{vertex degree}
\begin{equation}
 d_{\mathcal{G}}(x) := |\{y \in\zd : [x,y] \in \mathcal{E}\}|
\end{equation}
of $x\in\zd$ counts the number of edges in $\mathcal{G}$ that share the vertex
$x$ as an endpoint. Here, $|\Lambda|$ denotes the cardinality of a subset
$\Lambda \subset\zd$, and we use the convention $|\varnothing| =0$ for the
empty set.  A graph is called \emph{finite}, if $|\mathcal{V}| < \infty$.

A given graph $\mathcal{G}$ consists of finitely or infinitely many
\emph{clusters} $\mathcal{C}_{j}$, $j=1,2,\ldots, J \le \infty$, which
are the \emph{maximally connected subgraphs} of $\mathcal{G}$. More
precisely, $\mathcal{C}_{j} := (\mathcal{V}_{j}, \mathcal{E}_{j})$ is
a connected subgraph of $\mathcal{G}$, if $\varnothing \neq
\mathcal{V}_{j} \subseteq \mathcal{V}$, $\mathcal{E}_{j} \subseteq
\mathcal{E}$ and if for every pair $x,y \in\mathcal{V}_{j}$ with $x\neq y$
there exists $K\in\mathbb{N}$ and $x_{k}\in\mathcal{V}_{j}$,
$k=0,1,2,\ldots,K$, such that $x_{0}:=x$, $x_{K}:=y$ and
$[x_{k-1,},x_{k}] \in\mathcal{E}_{j}$ for all $k=1,2, \ldots, K$. A
connected subgraph $\mathcal{C}_{j}$ of $\mathcal{G}$ is maximal, and
hence a cluster, if for every connected subgraph $\mathcal{C}' :=
(\mathcal{V}', \mathcal{E}')$ of $\mathcal{G}$ obeying $\mathcal{V}'
\supseteq \mathcal{V}_{j}$ and $\mathcal{E}' \supseteq
\mathcal{E}_{j}$ one has $\mathcal{C}' = \mathcal{C}_{j}$. This
definition of clusters also includes \emph{isolated vertices}, in
which case $\mathcal{C}_{j} = (\{x\},\varnothing)$ for the
corresponding $x\in\mathcal{V}$ with $d_{\mathcal{G}}(x) =0$.
Apparently, the decomposition of $\mathcal{G}$ into its clusters is
unique -- apart from enumeration.

Next, we consider the probability space $\Omega =\{0,1\}^{\mathbb{E}^{d}}$,
which is endowed with the usual product sigma-algebra, generated by finite
cylinder sets, and equipped with a product probability measure $\mathbb{P}$.
Elementary events in $\Omega$ are sequences of the form $\omega \equiv
(\omega_{[x,y]})_{[x,y] \in \mathbb{E}^{d}}$, and we assume their entries to
be independently and identically distributed according to a Bernoulli law
\begin{equation}
  \mathbb{P} (\omega_{[x,y]} =1 ) =p
\end{equation}
with parameter $p \in ]0,1[$, the \emph{bond probability}. To a given
$\omega\in\Omega$ we associate the edge set
\begin{equation}
    \mathcal{E}^{(\omega)} := \bigl\{ [x,y] \in \mathbb{E}^{d} :
    \omega_{[x,y]} =1 
    \bigr\}\,.
\end{equation}

\begin{definition} 
  \label{bondgraph}
  The mapping $\Omega \ni \omega \mapsto \mathcal{G}^{(\omega)} :=
  (\mathbb{Z}^{d}, \mathcal{E}^{(\omega)})$ with values in the set of
  subgraphs of $\mathbb{L}^{d}$ is called \emph{bond-percolation graph in
  $\zd$}. 
\end{definition}
The most basic properties of bond-percolation graphs are recalled in

\begin{proposition}
  \label{percolation}
  For $d \ge 2$ there exists $p_{c}\in ]0,1[$, depending on $d$, such that
  \begin{nummer}
  \item \label{nonperc}
    for every $p \in ]0,p_{c}[$, the non-percolating phase, one has
    \begin{equation}
      \mathbb{P}\biggl\{\omega\in\Omega :
      ~~\parbox{4cm}{$\mathcal{G}^{(\omega)}$ \upshape
        consists of $\infty$--many\\ clusters, which are all
        finite} \biggr\} =1 \,,
    \end{equation}
  \item for every $p\in ]p_{c},1[$, the percolating phase, one has
    \begin{equation}
      \mathbb{P}\biggl\{\omega\in\Omega :
      ~~\parbox{5cm}{$\mathcal{G}^{(\omega)}$ \upshape
        consists of exactly one infinite\\ cluster and $\infty$--many finite
        clusters} \biggr\} =1 \,.
    \end{equation}
  \end{nummer}
\end{proposition}

\begin{remarks}
  \begin{nummer}
  \item 
    The proposition collects results from Thms.~1.10, 1.11, 4.2 and~8.1 in
    \cite{Gri99}, which were mainly obtained by Hammersley in the late
    fifties. The uniqueness of the infinite cluster, however, was only proven
    thirty years later by Aizenman, Kesten and Newman, see \cite{Gri99}.
  \item \label{oneDperc}
    In the one-dimensional situation, $d=1$, one has $p_{c}=1$ and
    part~\itemref{nonperc} of the proposition remains true.
  \end{nummer}
\end{remarks}

%
\subsection{Graph Laplacians}  \label{graphlap}
%

The subsequent definition introduces Laplacian-type operators associated with
an arbitrary subgraph of the integer lattice. The particularisation to
operators on bond-percolation graphs follows at the end of this subsection.

For a given subset $\Lambda\subseteq\zd$ let $\ell^{2}(\Lambda)$ denote the
Hilbert space of complex-valued, square-summable sequences that are indexed by
$\Lambda$.

\begin{definition}
  Given any subgraph $\mathcal{G} = (\mathcal{V}, \mathcal{E})$ of \,
  $\mathbb{L}^{d}$ with $\mathcal{V} \neq \varnothing$, we introduce the
  following bounded and self-adjoint linear operators on
  $\ell^{2}(\mathcal{V})$.
  \begin{nummer}
  \item The \emph{degree operator} $D(\mathcal{G})$ is defined as the
    multiplication operator with the vertex-degree function $d_{\mathcal{G}}:
    \mathcal{V} \rightarrow \mathbb{N} \cup \{0\}$, $x \mapsto
    d_{\mathcal{G}}(x)$, 
    that is,
    \begin{equation}
      [D(\mathcal{G})\varphi](x)  := d_{\mathcal{G}}(x) \,\varphi(x)
    \end{equation}
    for all $\varphi\in\ell^{2}(\mathcal{V})$ and all $x\in\mathcal{V}$.
  \item The \emph{adjacency operator} $A(\mathcal{G})$ is defined through its
    action 
    \begin{equation}
      [A(\mathcal{G})\varphi](x) := \sum_{y \in \mathcal{V}:\; [x,y] \in
      \mathcal{E}} \varphi(y)
    \end{equation}
    for all $\varphi\in\ell^{2}(\mathcal{V})$ and all $x\in\mathcal{V}$. Here,
    we use the convention $ \sum_{y\in\varnothing} \varphi(y) =0$ for the empty
    sum. 
  \item The \emph{Neumann Laplacian} is defined by
    \begin{equation}
      \DN(\mathcal{G}) := D(\mathcal{G}) - A(\mathcal{G})\,.  
    \end{equation}
  \item The \emph{Pseudo-Dirichlet Laplacian} is defined by
    \begin{equation}
      \label{DTdef}
      \DDT(\mathcal{G}) := \bigl(2d \one - D(\mathcal{G})\bigr) +
      \DN(\mathcal{G})   =  2d \one  - A(\mathcal{G})\,,
    \end{equation}
    where $\one \equiv \one_{\mathcal{V}}$ stands for the identity operator on
    $\ell^{2}(\mathcal{V})$.
  \item The \emph{Dirichlet Laplacian} is defined by
    \begin{equation}
      \label{Ddef}
      \DD(\mathcal{G}) := 2 \bigl(2d \one - D(\mathcal{G})\bigr) +
      \DN(\mathcal{G}) = 2d\one + \bigl(2d \one - D(\mathcal{G})\bigr) -
      A(\mathcal{G}) . 
    \end{equation}
  \end{nummer}
\end{definition}

\begin{remarks}
  \begin{nummer}
  \item The asserted boundedness and self-adjointness of the Laplacians
    $\DX(\mathcal{G})$, $\XinXset$, follow from the corresponding properties
    of $D(\mathcal{G})$ and $A(\mathcal{G})$. Indeed, since $0\le
    d_{\mathcal{G}}(x)\le 2d$ for all $x\in\mathcal{V}$, it is clear that
    $D(\mathcal{G})$ is self-adjoint and obeys $0 \le D(\mathcal{G}) \le
    2d\one$ in the sense of quadratic forms. The operator $A(\mathcal{G})$ is
    symmetric, because
    \begin{equation}
      \label{Aform}
      \langle \psi, A(\mathcal{G})\varphi\rangle = 2 \sum_{[x,y]
        \in\mathcal{E}}  \psi^{*}(x)\,\varphi(y) 
    \end{equation}
    for all $\psi, \varphi\in\ell^{2}(\mathcal{V})$, where
    $\langle\psi,\varphi\rangle:= \sum_{x\in\mathcal{V}}
    \psi^{*}(x)\,\varphi(x)$ denotes the standard Hilbert-space scalar product
    on $\ell^{2}(\mathcal{V})$. The factor 2 in \eqref{Aform} reflects that
    the sum is over \emph{un}ordered pairs.  Moreover, applying the
    Cauchy--Schwarz inequality to \eqref{Aform}, yields the upper bound $2d$
    for the usual operator norm of $A(\mathcal{G})$, and self-adjointness
    follows from symmetry and boundedness.
  \item The Neumann Laplacian $\DN(\mathcal{G})$ is called \emph{graph
      Laplacian} or \emph{combinatorial Laplacian} in spectral graph theory,
    where it is commonly studied in various forms
    \cite{Moh91,CvDo95,Chu97,Col98}.
  \item \label{nullspace}
    The quadratic form
    \begin{equation}
      \label{quadsum}
      \langle\varphi, \DN(\mathcal{G}) \varphi \rangle =
      \sum_{[x,y]\in\mathcal{E}} 
      \bigl| \varphi(x) -\varphi(y)\bigr|^{2}\,,
      \qquad\quad\varphi\in\ell^{2}(\mathcal{V})\,,  
    \end{equation}
    for the Neumann Laplacian reveals that 
    \begin{equation}
      \label{Npos}
      \DN(\mathcal{G}) \ge 0\,. 
    \end{equation}
    Thus, a necessary and sufficient condition for
    $\varphi\in\ell^{2}(\mathcal{V})$ to belong 
    to the zero-eigenspace of $\DN(\mathcal{G})$ is that $\varphi$ stays
    constant within each of the finite clusters of $\mathcal{G}$ (separately).
    Consequently, each finite cluster of $\mathcal{G}$ contributes
    exactly one zero eigenvalue to $\DN(\mathcal{G})$. In contrast, zero is
    not an eigenvalue of $\DDT(\mathcal{G})$ and $\DD(\mathcal{G})$.
  \item \label{clusterrem} Let $\XinXset$ and let $\mathcal{C}_{j} :=
    (\mathcal{V}_{j}, \mathcal{E}_{j})$, $j=1,\ldots J \le\infty$ denote the
    clusters a graph $\mathcal{G}:=(\mathcal{V},\mathcal{E})$ is composed of.
    Then $\DX(\mathcal{G})$ is block-diagonal with respect to the clusters,
    \begin{equation}
      \label{decomp}
      \DX(\mathcal{G}) = \ooplus_{j=1}^{J}
      \DX(\mathcal{C}_{j}) 
    \end{equation}
    on $\ell^{2}(\mathcal{V})$. We note that if $\mathcal{C}_{j}$
    corresponds to an isolated vertex, then $\DX(\mathcal{C}_{j})$
    acts as multiplication by $\gamma_{\mathrm{X}}$ on the
    one-dimensional subspace $\ell^{2}(\mathcal{V}_{j})$, where
    $\gamma_{\mathrm{N}} := 0$, $\gamma_{\,\;\widetilde{\mathrm{D}}}
    := 2d$, respectively $\gamma_{\;\mathrm{D}} := 4d$.
  \item \label{relate}
    The Neumann and the Dirichlet Laplacian are related to each other. To
    see this we define a unitary involution $U=U^{-1}=U^{*}$ on
    $\ell^{2}(\mathcal{V})$ by 
    setting
    \begin{equation}
      (U\varphi)(x) := (-1)^{\sum_{\nu =1}^{d}|x_{\nu}|}
      \varphi(x) 
    \end{equation}
    for all $x\in\mathcal{V}$ and all $\varphi\in\ell^{2}(\mathcal{V})$. This
    involution commutes with $D(\mathcal{G})$ and anti-commutes with
    $A(\mathcal{G})$ so that $D(\mathcal{G}) = U D(\mathcal{G}) U$ and 
    \begin{equation}
      \label{UAU}
      A(\mathcal{G}) = -U A(\mathcal{G}) U\,.
    \end{equation}
    Hence, we infer the relation
    \begin{equation}
      \label{DiNeu}
      \DD(\mathcal{G}) = 4d\one -U \DN(\mathcal{G}) U\,.
    \end{equation}
    Combining \eqref{DiNeu} with \eqref{Npos}, \eqref{DTdef} and~\eqref{Ddef},
    we arrive at the chain of inequalities
    \begin{equation} 
      \label{chain}
      0 \le \DN(\mathcal{G}) \le \DDT(\mathcal{G}) \le \DD(\mathcal{G}) \le
      4d\one \,.
    \end{equation}
  \item Our terminology of the Laplacians is motivated by Simon \cite{Sim85}.
    Divide a graph $\mathcal{G}=(\mathcal{V}, \mathcal{E})$ into two subgraphs
    $\mathcal{G}_{k}=(\mathcal{V}_{k}, \mathcal{E}_{k})$, $k=1,2$, such that
    $\mathcal{V}_{1}\cap\mathcal{V}_{2} = \varnothing$ and
    $\mathcal{V}_{1}\cup\mathcal{V}_{2} = \mathcal{V}$. Then one gets super-,
    respectively subadditive behaviour
    \begin{equation}
      \label{subsuper}
      \begin{split}
        \DN(\mathcal{G}) &\ge \DN(\mathcal{G}_{1}) \ooplus \DN(\mathcal{G}_{2})
        \,,\\ 
        \DD(\mathcal{G}) &\le \DD(\mathcal{G}_{1}) \ooplus \DD(\mathcal{G}_{2})
      \end{split}
    \end{equation}
    as a consequence of \eqref{quadsum} and \eqref{DiNeu}. Thus, introducing a
    separating boundary surface lowers Neumann eigenvalues and raises
    Dirichlet eigenvalues -- in analogy to the well-known behaviour of
    Laplacian eigenvalues of regions in continuous space, see e.g.\ Prop.~4 in
    Chap.\ XIII.15 of \cite{ReSi78}. In contrast, the eigenvalues of the
    Pseudo-Dirichlet Laplacian $\DDT(\mathcal{G})$ behave indifferently with
    respect to this procedure.  Though, $\DDT(\mathcal{G})$ is commonly termed
    a Dirichlet Laplacian in the literature.
  \end{nummer}
\end{remarks}

Next, we associate Laplacians to the bond-percolation graphs of
Definition\ref{bondgraph}.
\begin{definition}
  The mapping $\DX : \Omega \ni \omega \mapsto \DX^{(\omega)}
  := \DX(\mathcal{G}^{(\omega)})$ with values in the
  bounded, self-adjoint operators on $\l2zd$
  is called \emph{Neumann or Pseudo-Dirichlet or Dirichlet
    Laplacian on bond-percolation graphs in $\zd$}, depending on whether
  \emph{X} stands for \emph{N} or $\widetilde{\mathrm{D}}$ or \emph{D}.
\end{definition}

%
\subsection{Results} \label{results}
%

To begin with we summarise the most basic spectral properties of the Laplacian
on bond-percolation graphs in

\begin{lemma}
  \label{ergodic}
  Fix $\XinXset$ and $p\in ]0,1[$. Then
  \begin{nummer}
  \item \label{genergodic}
    the random operator $\DX$ is ergodic with respect to
    $\zd$-translations. 
  \item \label{specset} its spectrum is $\mathbb{P}$-almost surely non random,
    more precisely, it is given by $\spec (\DX) = [0,4d] \quad \mathbb{P}$-almost surely.
  \item \label{speccomp} the components in the Lebesgue decomposition
    of the spectrum are also $\mathbb{P}$-almost surely non random. For every
    $\varkappa \in \{\mathrm{pp}, \mathrm{sc}, \mathrm{ac}\}$ there exists a
    closed subset $\Sigma_{\mathrm{X}}^{(\varkappa)} \subset\mathbb{R}$ such
    that $\spec_{\varkappa} (\DX) = \Sigma_{\mathrm{X}}^{(\varkappa)} \quad
    \mathbb{P}$-almost surely.
  \item \label{specnonperc}
    in the non-percolating phase, $p \in ]0,p_{c}[$, the spectrum of $\DX$
    is $\mathbb{P}$-almost surely only a dense pure-point spectrum with
    infinitely degenerate eigenvalues. The dense set of eigenvalues is also
    non random $\mathbb{P}$-almost surely.
  \end{nummer}
\end{lemma}

\begin{remarks}
  \begin{nummer}
  \item The lemma is proven in Section~\ref{proofs}.
  \item Part~\itemref{specset} implies that the discrete spectrum of $\DX$ is
    $\mathbb{P}$-almost surely empty.
  \item \label{infcont} As compared to the non-percolating phase considered in
    part~\itemref{specnonperc}, there are additional spectral contributions
    from the percolating cluster if $p \in]p_{c}, 1[$. Among others, the
    percolating cluster contributes also infinitely degenerate,
    $\mathbb{P}$-almost surely non-random eigenvalues corresponding to
    compactly supported eigenfunctions.  This can be established with the same
    mirror techniques as it was done for related models on site-percolation
    graphs \cite{ChCh86,Ves05a,Ves05b}.  Non-rigorous arguments
    \cite{KiEg72,ShAh82,BiMo99} suggest the existence of continuous spectrum
    if $p$ lies above the ``quantum-percolation threshold'' $p_{q} >p_{c}$.
  \end{nummer}
\end{remarks}

We proceed with the existence and self-averaging of the integrated density of
states of $\DX$. To this end let $\delta_{x} \in \l2zd$ be the sequence which
is concentrated at the point $x\in\zd$, i.e.\ $\delta_{x}(x) :=1$ and
$\delta_{x}(y) := 0$ for all $y \neq x \in\zd$. Moreover, $\Theta$ stands for
the Heaviside unit-step function, which we choose to be right continuous,
viz.\ $\Theta(E) := 0$ for all real $E<0$ and $\Theta (E) := 1$ for all real $E
\ge0$.

\begin{definition}
  \label{Ndef}
  For every $p \in ]0,1[$ and every $\XinXset$ we call the function 
  \begin{equation}
    \NX : \mathbb{R} \ni E \mapsto \NX(E) :=
    \int_{\Omega}\!\mathbb{P}(\d\omega)  \; \langle\delta_{0} ,  
    \Theta\bigl(E - \DX^{(\omega)}\bigr) \delta_{0}\rangle 
  \end{equation}
  with values in the interval
  $[0,1]$ the \emph{integrated density of states of} $\DX$.
\end{definition}

\begin{remarks}
  \begin{nummer}
  \item Thanks to the ergodicity of $\DX$ with respect to $\zd$-translations,
    one can replace $\delta_{0}$ by $\delta_{x}$ with some arbitrary $x\in\zd$
    in Definition~\ref{Ndef} without changing the result.
  \item The integrated density of states $\NX$ is the right-continuous
    distribution function of a probability measure on $\mathbb{R}$. The set of
    its growth points coincides with the $\mathbb{P}$-almost-sure spectrum
    $[0,4d]$ of $\DX$.
  \item For $p < p_{c}$ the growth points of $\NX$ form a dense countable set,
    where $\NX$ is discontinuous. These jumps in $\NX$ are due to the
    infinitely degenerate eigenvalues of $\DX$, which arise solely from the
    finite clusters, cf.\ Lemma~\ref{specnonperc}. For $p>p_{c}$ there are
    also contributions to the jumps that arise from the percolating
    cluster. In addition, the set of growth points of $\NX$ should not be
    restricted to discontinuities for $p>p_{c}$, cf.\ Remark~\ref{infcont}.
  \item 
    Eqs.\ \eqref{UAU} and \eqref{DiNeu} imply the symmetries
    \begin{equation}
      \label{bcrel}
      \begin{split}
        \NDT (E) &= 1 - \lim_{\varepsilon\uparrow 4d -E} \NDT(\varepsilon)\,, \\
        N_{\mathrm{D(N)}}(E) &= 1 - \lim_{\varepsilon\uparrow 4d -E}
        N_{\mathrm{N(D)}} (\varepsilon)
      \end{split}
    \end{equation}
    of the integrated densities of states for all $E\in\mathbb{R}$.
  \end{nummer}
\end{remarks}
Definition~\ref{Ndef} of the integrated density of states coincides with the
usual one in terms of a macroscopic limit. To make this statement precise, we
have to introduce restrictions of $\DX$ to finite volume.

\begin{definition}
  Let $\mathcal{G}= (\mathcal{V}, \mathcal{E})$ be a subgraph of
  $\mathbb{L}^{d}$ and consider a subset $\Lambda\subseteq\zd$. 
  \begin{nummer}
  \item The graph $\mathcal{G}_{\Lambda} := (\mathcal{V}_{\Lambda},
    \mathcal{E}_{\Lambda})$ with $\mathcal{V}_{\Lambda} := \mathcal{V} \cap
    \Lambda$ and $\mathcal{E}_{\Lambda} := \{[x,y] \in \mathcal{E} : x,y \in
    \mathcal{V}_{\Lambda}\}$ is called the \emph{restriction of $\mathcal{G}$
      to $\Lambda$}. In particular, $\mathcal{G}^{(\omega)}_{\Lambda} =
    (\Lambda, \mathcal{E}_{\Lambda}^{(\omega)})$ is the restriction to
    $\Lambda$ of a realisation $\mathcal{G}^{(\omega)} =(\mathbb{Z}^{d},
    \mathcal{E}^{(\omega)})$ of the bond-percolation graph.
  \item For $\XinXset$ we define the restriction $\Delta_{\mathrm{X},
      \Lambda}$ of the Laplacian $\DX$ to $\ell^{2}(\Lambda)$ as the random
    operator with realisations $\Delta_{\mathrm{X}, \Lambda}^{(\omega)} :=
    \Delta_{\mathrm{X}}(\mathcal{G}^{(\omega)}_{\Lambda})$ for all $\omega \in
    \Omega$. 
  \end{nummer}
\end{definition}

\begin{lemma}
  \label{vollimit}
  Given $p \in ]0,1[$ and $\XinXset$, there exists a set $\Omega'
  \subset \Omega$ of full probability, $\mathbb{P}(\Omega') =1$, such that
  \begin{equation}
    \label{Nlimit}
    \NX(E) = \lim_{\Lambda \uparrow\zd} \left[\frac{1}{|\Lambda|} \;
    \tr\nolimits^{\phantom{y}}_{\ell^{2}(\Lambda)} \Theta \bigl(E-
    \Delta^{(\omega)}_{\mathrm{X}, \Lambda}\bigr) \right]
  \end{equation}  
  holds for all $\omega \in \Omega'$ and all $E\in\mathbb{R}\,$, except for
  the (at most countably many) discontinuity points of $\NX$.
\end{lemma}

\begin{remarks}
  \begin{nummer}
  \item \label{limitrem}
    As to the limit $\Lambda \uparrow\zd$, we think of a sequence of 
    cubes centred at the origin whose edge lengths tend to infinity. But there
    exist more general sequences of expanding regions in $\zd$ for which the
    lemma remains true.
  \item The proof of Lemma~\ref{vollimit} for $\mathrm{X} \in
    \{\mathrm{N},\mathrm{D}\}$ follows from the Ackoglu--Krengel superergodic
    theorem on account of \eqref{subsuper}, see Thm.~VI.1.7 in \cite{CaLa90}
    and the discussion after Eq.\ (VI.16) there.  For $\mathrm{X} =
    \widetilde{\mathrm{D}}$, the proof follows from Lemma~4.5 in
    \cite{PaFi92}, which establishes weak convergence of the associated
    density-of-states probability measures, and Thm.~30.13 in \cite{Bau92}.
  \item \label{convergeall} 
    The arguments in Sec.~6 of \cite{Ves05b} show that the convergence in
    \eqref{Nlimit} holds whenever $E$ is an algebraic number, that is the root
    of a polynomial with integer coefficients. Hence, the convergence
    \eqref{Nlimit} may even hold at discontinuity points of $\NX$.  In
    particular, for $p <p_{c}$ it holds for all $E\in\mathbb{R}$.
  \end{nummer}
\end{remarks}

The central result of this paper is

\begin{theorem}
  \label{main}
  Let $d\in\mathbb{N}$ and assume $p \in ]0,p_{c}[$. Then the integrated
  density of states $\NX$ of the Laplacian $\DX$ on bond-percolation graphs
  in $\zd$ exhibits Lifshits tails at both the lower
  spectral edge 
  \begin{equation}
    \label{lowertail}
    \lim_{E\downarrow 0}\;\frac{\ln\bigl| \ln [\NX(E) -\NX(0)]\bigr|}{\ln E} =
    \left\{ \begin{array}{r}  -1/2 \\  -d/2  \end{array}
    \right. \quad \mathit{for} \quad
    \begin{array}{l}
      \mathrm{X} = \mathrm{N}\,,\\
        \mathrm{X}= \widetilde{\mathrm{D}}, \mathrm{D}
    \end{array}
  \end{equation}
  and at the upper spectral edge
  \begin{equation}
    \label{uppertail}
    \lim_{E \uparrow 4d}\;\frac{\ln\bigl| \ln [\NX^{-}(4d) - \NX(E)]
    \bigr|}{\ln (4d -E)} =  
    \left\{ \begin{array}{r}  -1/2 \\  -d/2  \end{array}
    \right. \quad \mathit{for} \quad
    \begin{array}{l}
      \mathrm{X} = \mathrm{D}\,,\\
      \mathrm{X}= \mathrm{N}, \widetilde{\mathrm{D}}\,,
    \end{array}
  \end{equation}
  where $\NX^{-}(4d) := \lim_{E\uparrow 4d} \NX(E)$. 
\end{theorem}

\begin{remarks}
  \begin{nummer}
  \item The Lifshits tails at the upper spectral edge are related to the ones
    at the lower spectral edge due to the symmetries \eqref{bcrel}.
  \item \label{constants} 
    Remark \ref{nullspace}, the symmetries
    \eqref{bcrel}, Lemma~\ref{vollimit} and Remark \ref{convergeall} imply the
    values
    \begin{equation}
      \label{thmconstants}
      \begin{split}
        &\NDT(0) = \ND(0) =0, \qquad \qquad \NN^{-}(4d) = \NDT^{-}(4d) =1,
        \\[.5ex]
        &1 - \ND^{-}(4d) = \NN(0) = \lim_{\Lambda \uparrow\zd}
        \frac{\tr\nolimits^{\phantom{y}}_{\ell^{2} (\Lambda)} \Theta \bigl(
          -\Delta^{(\omega)}_{\mathrm{N}, \Lambda}\bigr) }{|\Lambda|} =
        \kappa(p) 
      \end{split}
    \end{equation}
    for the constants in Theorem~\ref{main}. Here, $\kappa(p) $ is the
    mean number density of clusters, see e.g.\
    Chap.~4 in \cite{Gri99}.  Thanks to the right-continuity of the
    Heaviside function, the operator $\Theta \bigl(
    -\Delta^{(\omega)}_{\mathrm{N}, \Lambda}\bigr)$ is nothing but the
    projector onto the null space of $\Delta^{(\omega)}_{\mathrm{N},
      \Lambda}$.
  \item The Lifshits tails for $\NN$ at the lower spectral edge -- and hence
    the one for $\ND$ at the upper spectral edge -- is determined by the
    linear clusters of bond-percolation graphs. This explains why the
    associated Lifshits exponent $-1/2$ is not affected by the spatial
    dimension $d$.  Technically, this relies on a Cheeger inequality
    \cite{Col98} for the second-lowest Neumann eigenvalue of a connected
    graph.
  \item If $d \ge2$, then all other Lifshits tails of the theorem are
    determined by the most condensed clusters of bond-percolation graphs, like
    cubic clusters (see Remark~\ref{completecube} below for their
    definition), as they maximise the mean vertex degree among all clusters
    with a given number of vertices. In the proof of the theorem this will
    follow from a Faber-Krahn inequality \cite{ChGr00} for the lowest
    \mbox{(Pseudo-)} Dirichlet eigenvalue of a connected graph. In contrast,
    for $d=1$ there are no other clusters than linear ones, and the Lifshits
    exponent cannot discriminate between different boundary conditions.
  \item For \emph{site}-percolation graphs, a stronger statement than
    \eqref{lowertail} is known for the case $\mathrm{X} =
    \widetilde{\mathrm{D}}$, see \cite{BiKo01}. 
  \end{nummer}
\end{remarks}

%
\section{Proofs}  \label{proofs}
%

In this section we shall prove Lemma~\ref{ergodic} and Theorem~\ref{main}.

\begin{proof}[of Lemma~\ref{ergodic}]
  We follow the standard arguments laid down in \cite{Kir89,CaLa90,PaFi92}.
  The function $\Omega \ni \omega \mapsto \DX^{(\omega)}$, which takes on
  values in the set of bounded self-adjoint operators on $\l2zd$, is
  measurable, and the probability measure $\mathbb{P}$ is ergodic with respect
  to the group of translations $(\tau_{z})_{z \in\zd}$ on $\Omega$, which act
  as $\tau_{z}\omega := (\omega_{[x+z,y+z]})_{[x,y]\in\zd}$. Moreover, for
  every $z\in\zd$ let $T_{z}$ be the unitary translation operator on $\l2zd$,
  that is, $T_{z}\varphi (x) := \varphi(x-z)$ for all $\varphi\in\l2zd$ and
  all $x\in\zd$. The operator identity $\DX^{(\tau_{z}\omega)} = T^{-1}_{z}
  \DX^{(\omega)} T_{z}$ holds for all $z\in\zd$ and all $\omega\in\Omega$ and
  renders $\DX$ an ergodic random operator \cite{Kir89}, as claimed in
  part~\itemref{genergodic} of the lemma. Part~\itemref{speccomp} is now a
  consequence of the general theory of ergodic random operators
  \cite{Kir89,CaLa90,PaFi92}.
  
  As to part~\itemref{specset} it suffices to show the inclusion
  \begin{equation}
    \label{specincl}
    \spec (\DX^{(\omega)}) \supseteq [0,4d] \quad \mbox{for
    $\mathbb{P}$-almost every $\omega\in\Omega$} 
  \end{equation}
  for all $\XinXset$, because the opposite inclusion is already supplied by
  \eqref{chain}. To verify \eqref{specincl}, we define the event
  \begin{align}
    \label{bcset}
    \widetilde{\Omega} := \Big\{ \omega \in\Omega : \;  \mbox{~for every~}
      & l\in\mathbb{N} \mbox{~there exists a cube $\Lambda_{l}^{(\omega)}
      \subset \zd$}  \nonumber \\ 
    & \mbox{~with $l^{d}$ points such that~}
    \mathcal{G}^{(\omega)}_{\Lambda_{l}^{(\omega)}} =
    \mathbb{L}^{d}_{\Lambda_{l}^{(\omega)}}\Big\}\,.
  \end{align}
  Here, we say that a subset of $\zd$ is a cube with $l^{d}$ points (or,
  equivalently, with edges of length $l-1 \in\mathbb{N}$), if this subset is
  some translate of the $d$-fold Cartesian product $\{1, \ldots,
  l\}^{d}$. Colloquially speaking, the condition in \eqref{bcset} requires all
  bonds inside of $\Lambda_{l}^{(\omega)}$ to be present. 
  
  Now, fix an arbitrary $E\in [0,4d] = \spec( \Delta)$
  in the spectrum of the ordinary lattice Laplacian $\Delta =
  \DX(\mathbb{L}^{d})$. 
  Then, there exists a Weyl sequence $(\psi_{E,n})_{n \in\mathbb{N}} \subset
  \l2zd$ for $\Delta$, that is, $\|\psi_{E,n}\| :=
  \langle\psi_{E,n}, \psi_{E,n}\rangle^{1/2} = 1$ for all $n\in\mathbb{N}$ and
  \begin{equation}
    \label{weyl}
    \lim_{n\to\infty} \| (\Delta - E\one) \psi_{E,n}\| =0\,.
  \end{equation}
  We may also assume without loss of generality that the support
  $\mathrm{supp}\,\psi_{E,n}$ is compact for all $n\in\mathbb{N}$, since
  $\Delta$ is bounded. Furthermore, if $(\psi_{E,n})_{n
    \in\mathbb{N}}$ is such a Weyl sequence, then so is
  $(T_{z_{n}}\psi_{E,n})_{n\in\mathbb{N}}$ with arbitrary $z_{n} \in\zd$.
  Thus, given any $\omega\in\widetilde{\Omega}$ there exists a Weyl sequence
  $(\psi_{E,n}^{(\omega)})_{n\in\mathbb{N}}$ for $\Delta$ with
  the property that, loosely speaking, all the supports are contained well
  inside the cubes of \eqref{bcset}.  More precisely, we mean that given every
  $\omega\in\widetilde{\Omega}$ and every $n\in\mathbb{N}$ there must exist
  an integer $l_{n}^{(\omega)} >3$ and a cube
  $\Lambda_{l_{n}^{(\omega)}}^{(\omega)}$ from \eqref{bcset} such
  that $\min\bigl\{|x-y| : x\in \mathrm{supp}\, \psi_{E,n}^{(\omega)}, y \in
  \zd\setminus \Lambda_{l_{n}^{(\omega)}}^{(\omega)} \bigr\} > 1$. This yields
  \begin{equation}
    \| (\DX^{(\omega)} - E\one) \psi_{E,n}^{(\omega)}\| 
     = \| (\Delta -  E\one)  \psi_{E,n}^{(\omega)}\| 
  \end{equation}
  for all $n\in\mathbb{N}$ and all $\omega\in\widetilde{\Omega}$. Hence,
  $(\psi_{E,n}^{(\omega)})_{n\in\mathbb{N}}$ is also a Weyl sequence for
  $\DX^{(\omega)}$, and we have shown the inclusion in \eqref{specincl} for all
  $\omega\in\widetilde{\Omega}$. But $\mathbb{P}(\widetilde{\Omega}) =1$, as
  we shall argue now.
  
  For every given integer $l \ge 2$ let $(\Lambda_{l,\mu})_{\mu\in\mathbb{N}}
  \subset \zd$ be a sequence of cubes in $\zd$ with $l^{d}$ points such that
  $\Lambda_{l,\mu_{1}} \cap \Lambda_{l,\mu_{2}} = \varnothing$, whenever
  $\mu_{1} \neq \mu_{2}$. Then, the events $\Omega_{l,\mu} := \{
  \omega\in\Omega : \mathcal{G}^{(\omega)}_{\Lambda_{l,\mu}} =
  \mathbb{L}^{d}_{\Lambda_{l,\mu}}\}$ are pairwise statistically independent,
  and $\mathbb{P}(\Omega_{l,\mu}) >0$ does not depend on $\mu\in\mathbb{N}$.
  So the Borel--Cantelli lemma implies $\mathbb{P}(\Omega_{l}) =1$ for all
  integers $l \ge 2$, where $\Omega_{l} := \limsup_{\mu\to\infty}
  \Omega_{l,\mu}$.  The proof of part~\itemref{specset} is completed by noting
  that $\widetilde{\Omega} \supset \cap_{l-1\in\mathbb{N}} \Omega_{l}$.
  
  Finally, we turn to part~\itemref{specnonperc} and assume $p <
  p_{c}$.  We observe that, in $\mathbb{P}$-almost every realisation of a
  bond-percolation graph, the translates of any given finite cluster
  occur infinitely often. This follows from a Borel--Cantelli argument
  like the one in the previous paragraph. Hence, the block-diagonal
  structure \eqref{decomp} of $\DX$ implies that the set of
  eigenvalues of $\DX$ is $\mathbb{P}$-almost surely given by the
  union of the spectra of $\DX(\mathcal{C})$, where $\mathcal{C}$ runs
  through all possible finite clusters in $\mathbb{L}^{d}$. In
  particular, the set of eigenvalues is a non-random dense set and all
  eigenvalues are infinitely degenerate.  \qed
\end{proof}

The remaining part of this section concerns the proof of Theorem~\ref{main}.
It relies on deterministic upper and lower bounds
for small eigenvalues of clusters. The lower bounds are discrete versions
of well-known isoperimetric estimates for Laplacian eigenvalues on manifolds.

\begin{definition}
  For $\XinXset$ and a connected subgraph
  $\mathcal{G}:=(\mathcal{V},\mathcal{E})$ of $\mathbb{L}^{d}$ with
  $|\mathcal{V}| \ge 2$ vertices, let
  $\EX(\mathcal{G})$ denote the lowest non-zero eigenvalue of
  $\DX(\mathcal{G})$.
\end{definition}

\begin{proposition}[Cheeger inequality]
  \label{chlemma}
  Let $\mathcal{G} := (\mathcal{V},\mathcal{E})$ be a connected finite
  subgraph of $\mathbb{L}^{d}$ with $|\mathcal{V}| \ge 2$ vertices.
  Then its lowest non-zero Neumann eigenvalue obeys
  \begin{equation}
    \label{cheeger}
    \EN(\mathcal{G}) \ge \frac{[h_{\mathrm{Ch}}(\mathcal{G})]^{2}}{4d} \,.
  \end{equation}
  The quantity $h_{\mathrm{Ch}}(\mathcal{G}) := \min_{\mathcal{K}}
  |\partial\mathcal{K}| / |\mathcal{K}|$ is the Cheeger constant, where the
  minimum is taken over all subgraphs $\mathcal{K}$ of $\mathcal{G}$ whose
  vertex set $\mathcal{W}$ obeys $|\mathcal{W}| \le |\mathcal{V}|/2$. Here,
  $|\partial\mathcal{K}| := \{ [x,y] \in \mathcal{E} : x\in \mathcal{W}, y \in
  \mathcal{V} \setminus \mathcal{W} \}$ denotes the edge boundary of
  $\mathcal{K}$ in $\mathcal{G}$.
\end{proposition}

\begin{remarks}
  \begin{nummer}    
  \item Proposition~\ref{chlemma} just quotes a special case of a more
    general, well-known result in graph theory, see e.g.\ Thm.~3.1(2) in
    \cite{Col98}. 
  \item \label{lincluster} The simple lower bound $h_{\mathrm{Ch}}
    (\mathcal{G}) \ge 1/(|\mathcal{V}|/2)$ on the Cheeger constant yields 
    \begin{equation}
      \label{crudech}
      \EN(\mathcal{G}) \ge \frac{d^{-1}}{|\mathcal{V}|^{2}}\,.
    \end{equation}
    This bound produces asymptotically the correct $|\mathcal{V}|$-dependence
    as $|\mathcal{V}|\to\infty$, if $\mathcal{G}$ is a linear cluster
    $\mathcal{L}_{n}$, i.e.\ a connected subgraph of $\mathbb{L}^{d}$ having 2
    vertices with degree 1 and $n-2$ vertices with degree 2. For highly
    connected clusters, such as cubic clusters in $d>1$ dimensions
    (see Remark~\ref{completecube} below for their definition), the bound
    \eqref{crudech} is very crude as compared to \eqref{cheeger}.  Though,
    \eqref{crudech} will suffice for our purpose.
  \end{nummer}
\end{remarks}

The next lemma provides a Faber--Krahn inequality on graphs. In contrast to
Cheeger inequalities, such estimates for graphs have not been known for a long
time, see \cite{ChGr00} for a detailed exposition. Lemma~\ref{fklemma} adapts
a result from \cite{ChGr00}, which is proven there for more general graphs, to
the type of graph Laplacians we use here.

\begin{lemma}[Faber--Krahn inequality]
  \label{fklemma}
  Let $\mathcal{G} := (\mathcal{V},\mathcal{E})$ be a connected finite
  subgraph of $\mathbb{L}^{d}$ with $|\mathcal{V}| \ge 2$ vertices.
  Then its lowest Pseudo-Dirichlet eigenvalue obeys
  \begin{equation}
    \EDT(\mathcal{G}) \ge \frac{h_{\mathrm{FK}}}{|\mathcal{V}|^{2/d}}\,,
  \end{equation}
  where $h_{\mathrm{FK}} \in]0,\infty[$ is a constant that depends only on the
  spatial dimension $d$.
\end{lemma}

\begin{remark}
  \label{completecube}
  The Faber--Krahn inequality produces asymptotically the correct
  $|\mathcal{V}|$-dependence as $|\mathcal{V}|\to\infty$, if, for example,
  $\mathcal{G}$ is a \emph{cubic cluster} $\mathcal{Q}_{l}$, that is,
  $\mathcal{Q}_{l} = \mathbb{L}^{d}_{\Lambda_{l}}$ for some finite cube
  $\Lambda_{l}\subset\zd$ with $|\Lambda_{l}| = l^{d}$ points, i.e.\ edges of
  length $l-1 \in\mathbb{N}$.
\end{remark}

\begin{proof}[of Lemma~\ref{fklemma}]
  We reduce the assertion to a particular case of Prop.~7.1 and Cor.~6.4 in
  \cite{ChGr00} by choosing the weighted graph in Prop.~7.1 as
  $\mathbb{L}^{d}$ with unit weights on all bonds -- note that these results
  in \cite{ChGr00} extend to $d=1$. Given any $\Lambda\subset
  \zd$, this yields the inequality
  \begin{equation}
    \label{grigor}
    \lambda_{1}(\Lambda) := \inf_{0 \neq\varphi \in c_{0}(\Lambda)}
    \frac{\langle \varphi, \DN(\mathbb{L}^{d})\varphi\rangle}{2d \,
      \langle\varphi, \varphi\rangle} 
    \ge \frac{\beta_{d}^{2}}{2 |\Lambda|^{2/d}}\,.
  \end{equation}
  The constant $\beta_{d}\in]0,\infty[$ is the isoperimetric constant of
  Cor.~6.4 in \cite{ChGr00}, which is independent of $\Lambda\subset\zd$.
  Moreover, $c_{0}(\Lambda)$ stands for the $\l2zd$-subspace of real-valued
  sequences with support in $\Lambda$.  In order to check that the above
  definition of $\lambda_{1}(\Lambda)$ matches the one in \cite{ChGr00}, we
  refer to Sect.~5.5 of that paper.  The claim of the lemma follows from the
  estimate
  \begin{equation}
    \label{reduce}
    \EDT(\mathcal{G}) \ge 2d \,\lambda_{1}(\mathcal{V})\,,
  \end{equation}
  which we prove now. To this end we observe that the adjacency operator has
  only non-negative matrix elements $\langle\delta_{x},
  A(\mathcal{G})\delta_{y}\rangle$ 
  for all $x,y\in\mathcal{V}$ and that $A(\mathcal{G})$ is irreducible on
  $\ell^{2}(\mathcal{V})$ due
  to the connectedness of $\mathcal{G}$. Hence, it follows from the
  Perron--Frobenius theorem, see e.g.\ Thm.~2.1.4 in \cite{BePl79}, that the
  eigenvector of $\DDT(\mathcal{G}) = 2d\one
  -A(\mathcal{G})$ corresponding to the
  non-degenerate smallest eigenvalue $\EDT(\mathcal{G})$ can be chosen to have
  non-negative entries. This implies
   \begin{equation}
     \label{fkminmax}
     \EDT(\mathcal{G}) = \inf_{
       \begin{subarray}{c}
         0 \neq \phi \in \ell^{2}(\mathcal{V}) \\
         \phi(x) \ge 0 \mathrm{~for~all~} x\in\mathcal{V}
       \end{subarray}
     }
     \frac{\langle\phi,\DDT(\mathcal{G})\phi\rangle}{\langle\phi,
     \phi\rangle} \,.
   \end{equation}
   The proof of \eqref{reduce} is completed by noting that
   \begin{align}
     \langle\phi,\DDT(\mathcal{G})\phi\rangle & =
     \sum_{[x,y]\in\mathcal{E}} [\phi(x) - \phi(y)]^{2} +
     \sum_{x\in\mathcal{V}} [2d -
     d_{\mathcal{G}}(x)]\bigl(\phi(x)\bigr)^{2} \nonumber\\
     &= \sum_{[x,y]\in\mathbb{E}^{d}} [\varphi(x) - \varphi(y)]^{2} + 2
     \sum_{[x,y]\in\mathbb{E}^{d} \setminus \mathcal{E}} \varphi(x)
     \varphi(y) \nonumber\\
     &\ge \langle\varphi,\DN(\mathbb{L}^{d})\varphi\rangle
   \end{align}
   for all $\phi$ as in \eqref{fkminmax}, where 
$\varphi(x):= \Bigl\{\begin{array}{cl} \phi(x), & x \in\mathcal{V}\!, 
  \\ 0, & x\in\zd\setminus\mathcal{V}\!.\end{array}$ \quad\qed
\end{proof}

As a last ingredient for the proof of Theorem~\ref{main} we need some simple
upper estimates on $\EX(\mathcal{G})$ for special types of clusters. These
estimates are obtained from the minmax-principle. For our purpose it is only
important that they reproduce the asymptotically correct functional dependence
on the number of vertices for large clusters.

\begin{lemma}
  \begin{nummer}
  \item \label{eigenchain} The lowest non-zero Neumann eigenvalue for a linear
    cluster $\mathcal{L}_{n}$, which was defined in Remark~\ref{lincluster},
    obeys
    \begin{equation}
      \EN(\mathcal{L}_{n}) \le \frac{12}{n^{2}}
    \end{equation}
    for all $n\in\mathbb{N}$, $n\ge 2$. 
  \item \label{eigencube} The lowest Dirichlet eigenvalue for
    a cubic cluster $\mathcal{Q}_{l}$, which was defined in
    Remark~\ref{completecube}, obeys
    \begin{equation}
      \ED(\mathcal{Q}_{l}) \le \frac{27d}{l^{2}}
    \end{equation}
    for all $l \in\mathbb{N}$, $l\ge 2$.     
  \end{nummer}
\end{lemma}

\begin{proof}
  \begin{nummer}
    \item The minmax-principle yields the upper estimate 
      \begin{equation}
         \EN(\mathcal{L}_{n}) \le \frac{\sum_{j=1}^{n-1} (u_{j+1}
           -u_{j})^{2}}{\sum_{j=1}^{n} u_{j}^{2}}
      \end{equation}
      for every $(u_{1},\ldots, u_{n})\in\mathbb{R}^{n}$ subject to the
      orthogonality constraint $\sum_{j=1}^{n} u_{j} =0$. Choosing $u_{j} :=
      -j + (n+1)/2$ for $j\in\{1,\ldots,n\}$, which is in accordance with the
      orthogonality constraint, proves part~\itemref{eigenchain}.
    \item Appealing to the minmax-principle with a trial ``function'' that
    factorises with respect to the $d$ Cartesian directions, gives 
    \begin{equation}
      \ED(\mathcal{Q}_{l}) \le d\; \frac{ 2 u_{1}^{2} + 2 u_{l}^{2} + 
        \sum_{j=1}^{l-1} (u_{j+1} -u_{j})^{2}}{\sum_{j=1}^{l} u_{j}^{2}}
    \end{equation}
    for all $(u_{1},\ldots, u_{l})\in\mathbb{R}^{l}$. Now, we choose $u_{j} :=
    - | j - (l+1)/2| + (l-1)/2$ for all $j\in\{1,\ldots,l\}$ so that
    $u_{1}=u_{l} = 0$. If $l \ge 3$ is odd, an explicit calculation shows 
    \begin{equation}
       \ED(\mathcal{Q}_{l}) \le  \frac{12d}{l^{2} -2l +3} \le
       \frac{27d}{l^{2}}\,,
    \end{equation}
    while for an even integer $l \ge 2$ it yields
    \begin{equation}
       \ED(\mathcal{Q}_{l}) \le  \frac{12d}{l(l-1)} \le \frac{24d}{l^{2}}\,,
    \end{equation}
    and the lemma is proven. \qed
  \end{nummer}
\end{proof}

The next two lemmas provide the key estimates for Theorem~\ref{main}. While
the lower bounds in Lemma~\ref{lowerlemma} hold for all $p \in]0,1[$, the
upper bounds in Lemma~\ref{upperlemma} are restricted to the non-percolating
phase.

\begin{lemma}[Upper bounds]
   \label{upperlemma}
   Let $d\in\mathbb{N}$ and consider $p\in ]0,p_{c}[$. Then there exist
   constants $\alpha_{\mathrm{N}}^{+},\alpha_{\mathrm{D}}^{+} \in ]0,\infty[$
   such that
   \begin{align}
    \label{upperN}
    \NN(E) -\NN(0) & \le \exp\{ - \alpha_{\mathrm{N}}^{+} E^{-1/2}\}
    \qquad \mbox{for all~~} E\in]0,4d]\,, \\
    \label{upperD}
    \ND(E)   \le \NDT(E) & \le \exp\{ - \alpha_{\mathrm{D}}^{+} E^{-d/2}\}
    \qquad \mbox{for all~~} E \in ]0, 2d[.
   \end{align}
\end{lemma}

\begin{remark}
  It is only the right inequality in \eqref{upperD} whose validity is
  restricted to $E \in ]0, 2d[$. The proof below will show that $\ND(E) \le
  \exp\{ - \alpha_{\mathrm{D}}^{+} E^{-d/2}\}$ holds for all $E \in ]0, 4d[$.
\end{remark}

\begin{lemma}[Lower bounds]
  \label{lowerlemma}
  Let $d\in\mathbb{N}$ and consider $p\in ]0,1[$. Then there exist constants
  $\alpha_{\mathrm{N}}^{-},\alpha_{\mathrm{D}}^{-} \in ]0,\infty[$
  such that for every $E \in ]0,4d]$ one has
  \begin{align}
    \label{lowerN}
    \NN(E) -\NN(0) & \ge \exp\{ - \alpha_{\mathrm{N}}^{-} E^{-1/2}\} \,, \\ 
    \label{lowerD}
    \NDT(E) \ge \ND(E) & \ge \exp\{ - \alpha_{\mathrm{D}}^{-} E^{-d/2}\} \, .  
  \end{align}
\end{lemma}

\begin{proof}[of Theorem \ref{main}]
  Due to the symmetries \eqref{bcrel}, it suffices to prove the asserted
  Lifshits tails at the lower spectral edge. These follow from the estimates
  in Lemma~\ref{upperlemma} and Lemma~\ref{lowerlemma}, because after taking
  appropriate logarithms, the respective bounds coincide in the limit
  $E\downarrow 0$. \qed
\end{proof}

\begin{proof}[of Lemma~\ref{upperlemma}]
  Fix $E \in ]0,4d]$ subject to $E < \gamma_{\,\mathrm{X}}$ if $\mathrm{X}
  \in\{\widetilde{\mathrm{D}}, \mathrm{D}\}$.  The constants
  $\gamma_{\,\mathrm{X}}$ were defined in Remark~\ref{clusterrem}.
  Definition~\ref{Ndef} of the integrated density of states implies
  \begin{equation}
    \label{startlower}
    \NX(E) -\NX(0) = \int_{\Omega}\!\mathbb{P}(\d\omega)\; \bigl\langle 
    \delta_{0} , \bigl[\Theta\bigl(E - \DX^{(\omega)}\bigr) - \PX^{(\omega)}
    \bigr] \delta_{0}\bigr\rangle  \,,
  \end{equation}
  where $\PX := \Theta(-\DX)$ denotes the (random) projector onto the
  null space of $\DX$. Due to our assumptions on $E$ and the block-diagonal
  form \eqref{decomp} of $\DX$, the right-hand side of \eqref{startlower} is
  only different from zero if the origin is part of a cluster with at
  least two vertices. Let us call
  this event $\Omega_{0}$ and the corresponding cluster
  $\mathcal{C}_{0}^{(\omega)} := (\mathcal{V}_{0}^{(\omega)},
  \mathcal{E}_{0}^{(\omega)})$ for all $\omega \in\Omega_{0}$. Hence, we
  obtain
  \begin{align}
   \label{predistinct}
    \NX(E) -\NX(0) & = \int_{\Omega_{0}}\!\mathbb{P}(\d\omega)\; \bigl\langle 
    \delta_{0} , \bigl[\Theta\bigl(E - \DX(\mathcal{C}_{0}^{(\omega)})\bigr) -
    \PX(\mathcal{C}_{0}^{(\omega)}) \bigr] \delta_{0}\bigr\rangle \nonumber\\
    & \le  \int_{\Omega_{0}}\!\mathbb{P}(\d\omega)\; \Theta\bigl(E -
    \EX(\mathcal{C}_{0}^{(\omega)})\bigr) \bigl\langle  \,
    \delta_{0} , \bigl[ \one - \PX(\mathcal{C}_{0}^{(\omega)}) \bigr]
    \delta_{0}\bigr\rangle \nonumber\\ 
    & \le \mathbb{P}\bigl\{ \omega\in\Omega_{0} : 
    E \ge \EX(\mathcal{C}_{0}^{(\omega)}) \bigr\} \,.
  \end{align}
  Before we make a distinction of the two cases $\mathrm{X}=\mathrm{N}$ and
  $\mathrm{X}\in\{\widetilde{\mathrm{D}}, \mathrm{D}\}$ in order to apply
  the Cheeger, respectively the Faber--Krahn inequality, we recall from
  Proposition~\ref{percolation} (and Remark~\ref{oneDperc} for the case $d=1$)
  that the cluster $\mathcal{C}_{0}^{(\omega)}$ is finite for
  $\mathbb{P}$-almost all $\omega \in\Omega_{0}$, since we assume $p <p_{c}$.
  
  \emph{Neumann case.~} Applying the weakened version \eqref{crudech} of
  Cheeger's inequality to \eqref{predistinct}, yields the claim
  \begin{equation}
    \label{uNfinal}
    \NN(E) -\NN(0) \le \mathbb{P}\bigl\{ \omega\in\Omega_{0} : 
    |\mathcal{V}_{0}^{(\omega)}| \ge 1/(dE)^{1/2} \bigr\} 
    \le \exp\{ - d^{-1/2} \zeta(p) E^{-1/2}\} \,.
  \end{equation}
  The second inequality in \eqref{uNfinal} reflects the exponential decay of
  the cluster-size distribution in the non-percolating phase, see Thm.~6.75 in
  \cite{Gri99}. Here, $\zeta(p) >0$ is some finite constant for every
  $p\in]0,p_{c}[$, which depends only on $d$. Formally, Thm.~6.75 in
  \cite{Gri99} does not cover the one-dimensional situation $d=1$. But for
  $d=1$ the exponential decay of the cluster-size distribution follows from
  elementary combinatorics.

  \emph{(Pseudo-) Dirichlet case.~} 
  The inequalities \eqref{chain}, \eqref{predistinct}, the Faber--Krahn
  inequality of Lemma~\ref{fklemma} and the exponential decay of the
  cluster-size distribution yield
  \begin{equation}
    \ND(E)  \le \NDT(E) \le \mathbb{P}\bigl\{ \omega\in\Omega_{0} : 
    |\mathcal{V}_{0}^{(\omega)}| \ge (h_{\mathrm{FK}}/E)^{d/2} \bigr\} 
    \le \exp\{- h_{\mathrm{FK}}^{d/2} \zeta(p) E^{-d/2}\} \,.
  \end{equation}
  Here we have used that $\ND(0) = \NDT(0) = 0$, see Remark~\ref{constants}.
  \qed
\end{proof}

\begin{proof}[of Lemma~\ref{lowerlemma}]
  Lemma~\ref{vollimit}, the isotony of the right-hand side of \eqref{Nlimit}
  in $E$, the right-continuity of $\NX$ and \eqref{thmconstants} imply that 
  \begin{equation}
    \label{lowerstart}
    \NX(E) -\NX(0) \ge \limsup_{\Lambda\uparrow\zd} \frac{1}{|\Lambda|} \,
    \tr\nolimits_{\ell^{2}(\Lambda)} \bigl[ \Theta \bigl( E -
    \Delta_{\mathrm{X}, 
    \Lambda}^{(\omega)} \bigr) - P_{\mathrm{X}, \Lambda}^{(\omega)}\bigr]
  \end{equation}
  for \emph{all} $E>0$ and all $\omega\in\Omega'$. Concerning the limit
  in \eqref{lowerstart}, we think of a sequence of expanding cubes that are
  centred at the origin, cf. Remark~\ref{limitrem}. Due to the block-diagonal
  form \eqref{decomp} of $\DX$, which continues to hold for
  $\Delta_{\mathrm{X}, \Lambda}^{(\omega)}$ with respect to the decomposition
  of $\mathcal{G}_{\Lambda}^{(\omega)}$ into clusters
  $\mathcal{C}_{\Lambda,j}^{(\omega)} := (\mathcal{V}_{\Lambda,j}^{(\omega)},
  \mathcal{E}_{\Lambda,j}^{(\omega)})$, $j\in\{ 1, \ldots,
  J_{\Lambda}^{(\omega)} \}$, we get
  \begin{equation}
    \label{clustertrace}
    \NX(E) -\NX(0) \ge \limsup_{\Lambda\uparrow\zd} \frac{1}{|\Lambda|} \,
    \sum_{j=1}^{J_{\Lambda}^{(\omega)}}
    \tr\nolimits_{\ell^{2} (\mathcal{V}_{\Lambda,j}^{(\omega)})} 
    \bigl[ \Theta \bigl( E - \DX(\mathcal{C}_{\Lambda,j}^{(\omega)})\bigr)  -
    \PX(\mathcal{C}_{\Lambda,j}^{(\omega)}) \bigr]\,.
  \end{equation}
  At this point we make again a
  distinction of the cases $\mathrm{X}=\mathrm{N}$ and
  $\mathrm{X}\in\{\widetilde{\mathrm{D}}, \mathrm{D}\}$.
  
  \emph{Neumann case.~} Let $\mathfrak{L}_{n}$ denote the set of all
  linear clusters with $n \ge 2$ vertices in $\mathbb{L}^{d}$ and let
  $\Chi_{\!\mathfrak{L}_{n}}$ be the characteristic function of this set of 
  graphs. A crude lower bound on the $j$-sum in \eqref{clustertrace} for
  $\mathrm{X}=\mathrm{N}$ results from discarding all branched clusters, i.e.\
  those which are not linear, 
  \begin{align}
    \sum_{j=1}^{J_{\Lambda}^{(\omega)}} \sum_{n=2}^{\infty} &
    \Chi_{\!\mathfrak{L}_{n}}(\mathcal{C}_{\Lambda,j}^{(\omega)}) \;
    \tr\nolimits_{\ell^{2}(\mathcal{V}_{\Lambda,j}^{(\omega)})}
    \bigl[ \Theta \bigl( E - \DN(\mathcal{C}_{\Lambda,j}^{(\omega)})\bigr) -
    \PN(\mathcal{C}_{\Lambda,j}^{(\omega)}) \bigr] \nonumber\\
    & \ge  \sum_{n=2}^{\infty} \sum_{j=1}^{J_{\Lambda}^{(\omega)}}
    \Chi_{\!\mathfrak{L}_{n}}(\mathcal{C}_{\Lambda,j}^{(\omega)}) \,
    \Theta\bigl( E- \EN(\mathcal{C}_{\Lambda,j}^{(\omega)})\bigr) \nonumber\\
    & \ge \sum_{n=2}^{\infty} \Theta (E - 12/n^{2}) 
    \sum_{j=1}^{J_{\Lambda}^{(\omega)}}
    \Chi_{\!\mathfrak{L}_{n}}(\mathcal{C}_{\Lambda,j}^{(\omega)}) \,.
    \label{tracebound}
  \end{align}
  The first inequality in \eqref{tracebound} follows from restricting the
  trace to the spectral subspace corresponding to
  $\EN(\mathcal{C}_{\Lambda,j}^{(\omega)})$, the second inequality follows
  from the variational upper bound in Lemma~\ref{eigenchain}. Thus,
  \eqref{clustertrace} and \eqref{tracebound} yield
  \begin{equation}
    \label{prefatou}
    \NN(E) -\NN(0) \ge \limsup_{\Lambda \uparrow\zd} \sum_{n=2}^{\infty}
    \Theta (E - 12/n^{2}) \,  n^{-1} \, L_{n}^{(\omega)}(\Lambda)
  \end{equation}
  with 
  \begin{equation}
    L_{n}^{(\omega)}(\Lambda) := \frac{n}{|\Lambda|}
    \sum_{j=1}^{J_{\Lambda}^{(\omega)}} 
    \Chi_{\!\mathfrak{L}_{n}}(\mathcal{C}_{\Lambda,j}^{(\omega)}) 
    = \frac{1}{|\Lambda|} \; \bigl| \bigl\{ x \in \Lambda : 
    \mathcal{C}_{\Lambda}^{(\omega)}(x) \in\mathfrak{L}_{n} \bigr\}\bigr|
  \end{equation}
  being the number density of points in $\Lambda$ that are vertices of a
  cluster of type $\mathfrak{L}_{n}$. Here,
  $\mathcal{C}_{\Lambda}^{(\omega)}(x)$ denotes the cluster of
  $\mathcal{G}_{\Lambda}^{(\omega)}$ that contains $x\in\Lambda$. For
  a given $n\in\mathbb{N}$, $n\ge 2$, and a sufficiently large bounded
  cube $\Lambda \subset\zd$ with $|\Lambda|^{1/d} \ge 2n+1$, let us
  also define the number density
  \begin{equation}
    \widetilde{L}_{n}^{(\omega)} (\Lambda) := \frac{1}{|\Lambda|} \, \Bigl|
    \Bigl\{ x
    \in \Lambda : \min_{1 \le \nu\le d} \;\min_{y\in\zd\setminus \Lambda}
    |x_{\nu} - y_{\nu}| \ge n+1
    \mathrm{~and~~} 
    \mathcal{C}_{\Lambda}^{(\omega)}(x) \in\mathfrak{L}_{n} \Bigl\} \Bigr|    
  \end{equation}
  of vertices which are, in addition, sufficiently far away from the
  boundary of $\Lambda$. Clearly, one has
  \begin{equation}
    \label{nodiffer}
    \lim_{\Lambda\uparrow\zd} \bigl[ L_{n}^{(\omega)}(\Lambda) -
    \widetilde{L}_{n}^{(\omega)}(\Lambda)\bigr] =0
  \end{equation}
  for all $\omega\in\Omega$ and all $n\in\mathbb{N}$, $n\ge 2$, since the
  difference in the two quantities results from a surface effect. The vertices
  that count for $\widetilde{L}_{n}^{(\omega)}(\Lambda)$ are so far away from
  the boundary of $\Lambda$ that the clusters they belong to cannot grow when
  enlarging $\Lambda$. Hence,
  \begin{equation}
    |\Lambda_{1} \cup\Lambda_{2}| \, \widetilde{L}_{n}^{(\omega)}(\Lambda_{1}
    \cup\Lambda_{2}) \ge |\Lambda_{1}|\,
    \widetilde{L}_{n}^{(\omega)}(\Lambda_{1}) + |\Lambda_{2}|\,
    \widetilde{L}_{n}^{(\omega)}(\Lambda_{2})
  \end{equation}
  holds for all $\Lambda_{1}, \Lambda_{2} \subset\zd$ provided $\Lambda_{1}
  \cap \Lambda_{2} =\varnothing$. Thus, $L_{n}(\Lambda)$ defines a superergodic
  process and the Ackoglu--Krengel superergodic theorem, see e.g.\ Thm.~VI.1.7
  in \cite{CaLa90}, and \eqref{nodiffer} imply
  \begin{equation}
    \label{superergodic}
    \lim_{\Lambda\uparrow\zd} L_{n}^{(\omega)}(\Lambda) =
    \sup_{\Lambda \subset\zd} \int_{\Omega} \! \mathbb{P}(\d\omega') \;
    \widetilde{L}_{n}^{(\omega')}(\Lambda)
    = \mathbb{P} \bigl\{ \omega'\in\Omega_{0} :
    \mathcal{C}_{0}^{(\omega')} \in\mathfrak{L}_{n}\bigr\} 
  \end{equation}
  for all $n\in\mathbb{N}$, $n\ge 2$, and $\mathbb{P}$-almost all
  $\omega\in\Omega$. The event $\Omega_{0}$ and the random cluster
  $\mathcal{C}_{0}$ were defined above Eq.~\eqref{predistinct}. 
  
  Now, we neglect all terms in the $n$-sum in \eqref{prefatou} except for the
  one which corresponds to the biggest integer $n(E)$ obeying $n(E) <
  (12/E)^{1/2} +1$. From this we conclude together with \eqref{superergodic}
  that
  \begin{equation}
    \label{postfatou}
    \NN(E) -\NN(0) \ge  [n(E)]^{-1}\; \mathbb{P} \bigl\{
    \omega\in\Omega_{0} : \mathcal{C}_{0}^{(\omega)} 
    \in\mathfrak{L}_{n(E)}\bigr\} \,.
  \end{equation}
  Elementary combinatorics shows that the probability on the right-hand side
  of \eqref{postfatou} is bounded below by $\exp\{ -n(E) \, f(p)\}$, where
  $f(p) \in ]0,\infty [$ is a constant that depends only
  on $d$ for a given $p \in]0,1[$. This leads to the estimate
  \begin{equation}
    \NN(E) -\NN(0) \ge  \frac{\e^{- f(p)}}{(12/E)^{1/2} +1} \;
    \exp\bigl\{ - 12^{1/2} f(p) \,E^{-1/2}\bigr\} \,, 
  \end{equation}
  which can be cast into the form \eqref{lowerN} for $E\in ]0,4d]$.
  
  \emph{(Pseudo-) Dirichlet case.~} This case parallels exactly the
  previous one, except that here we retain cubic clusters instead of
  linear clusters. Let $\mathfrak{Q}_{l}$ denote the set of all cubic
  clusters in $\mathbb{L}^{d}$ with $l^{d}$ vertices, i.e.\ edges of length
  $l-1 \in\mathbb{N}$, let $\Chi_{\!\mathfrak{Q}_{l}}$ be the characteristic
  function of this set of graphs and define the number density
  $Q_{l}^{(\omega)}(\Lambda) := (l^{d}/|\Lambda|)
  \sum_{j=1}^{J_{\Lambda}^{(\omega)}}
  \Chi_{\!\mathfrak{Q}_{l}}(\mathcal{C}_{\Lambda,j}^{(\omega)})$ of points in
  $\Lambda$ that are vertices of such a cubic cluster (when
  restricted to $\Lambda$). Now, the r\^{o}le of $\mathfrak{L}_{n}$ in the
  previous case will be played by $\mathfrak{Q}_{l}$. Hence, the analogue of
  \eqref{prefatou} reads
  \begin{equation}
    \label{Dprefatou}
    \NDT(E) \ge \ND(E) \ge \limsup_{\Lambda \uparrow\zd}
    \sum_{l=2}^{\infty} \Theta \bigl(E - 27d/l^{2} \bigr)
    \,  l^{-d} \,  Q_{l}^{(\omega)}(\Lambda) \,,
  \end{equation}
  where we have used Lemma~\ref{eigencube} instead of
  Lemma~\ref{eigenchain}. The very same arguments that led to
  \eqref{superergodic} imply in the present context
  \begin{equation}
    \label{cubeprob}
    \lim_{\Lambda\uparrow\zd} Q_{l}^{(\omega)}(\Lambda) = 
    \mathbb{P} \bigl\{ \omega'\in\Omega_{0} :
    \mathcal{C}_{0}^{(\omega')} \in\mathfrak{Q}_{l}\bigr\} 
  \end{equation}
  for all $l\in\mathbb{N}$, $l\ge 2$, and $\mathbb{P}$-almost all
  $\omega\in\Omega$. By neglecting all terms in the $l$-sum in
  \eqref{Dprefatou} except for the one which corresponds to the biggest
  integer $l(E)$ obeying $l(E) < (27d/E)^{1/2} +1$, we conclude with 
  \eqref{cubeprob} that
  \begin{equation}
    \label{Dpostfatou}
     \NDT(E) \ge \ND(E) \ge [l(E)]^{-d} 
    \,\mathbb{P} \bigl\{ \omega\in\Omega_{0} : 
    \mathcal{C}_{0}^{(\omega)} \in\mathfrak{Q}_{l(E)}\bigr\} \,.
  \end{equation}
  Again, there is an elementary combinatorial lower bound $\exp\{ -[l(E)]^{d}
  \, g(p)\}$ for the probability in \eqref{Dpostfatou}, where $g(p) \in
  ]0,\infty [$ is a constant that depends only on $d$ for a given $p
  \in]0,1[$. So we arrive at
  \begin{equation}
    \NDT(E) \ge \ND(E) \ge [(27d/E)^{1/2} +1]^{-d} \exp\bigl\{ -
    [(27d/E)^{1/2} +1]^{d} g(p)\bigr\}\,,
  \end{equation}
  which can be cast into the form \eqref{lowerD} for $E\in ]0,4d]$. \qed
\end{proof}


\begin{acknowledgement}
  We are much obliged to Bernd Metzger and Peter Stollmann for
  pointing us to Ref.~\cite{ChGr00}. Our thanks go also to Ivan
  Veseli\'c for some helpful comments. PM received financial support
  by SFB/TR~12 of the Deutsche Forschungsgemeinschaft.
\end{acknowledgement}



\begin{thebibliography}{99}

\frenchspacing

\bibitem{Ant95}
  \au{P.}{Antal}%
  \ti{Enlargement of obstacles for the simple random walk}
  \z{Ann. Probab.}{23}{1061--1101}{1995}

\bibitem{Bau92}
  \au{H.}{Bauer}%
  \bti{Measure and integration theory}
  \pub*[German original: \pub{de Gruyter}{Berlin}{1992}]{de Gruyter}{Berlin}{2001} 
  
\bibitem{BePl79}
  \au{A.}{Berman}\et\au{R. J.}{Plemmons}%
  \bti{Nonnegative matrices in the mathematical sciences}
  \pub{Academic}{New York}{1979}
   
\bibitem{BiMo99}
  \au{G.}{Biroli}\et\au{R.}{Monasson}%
  \ti{A single defect approximation for localized states on random
    lattices}
  \z{J. Phys. A}{32}{L255--L261}{1999}

\bibitem{BiKo01}
  \au{M.}{Biskup}\et\au {W.}{K\"onig}%
  \ti{Long-time tails in the parabolic Anderson model with bounded potential}
  \z{Ann. Probab.}{29}{636--682}{2001}

\bibitem{BrRo88} 
  \au{A. J.}{Bray}\et\au{G. J.}{Rodgers}%
  \ti{Diffusion in a sparsely connected space: A model for glassy relaxation} 
  \z{Phys. Rev. B}{38}{11461--11470}{1988}

\bibitem{BrAs01} 
  \au{K.}{Broderix}, \au{T.}{Aspelmeier}, \au{A. K.}{Hartmann}\et
    \au{A.}{Zippelius}%
  \ti{Stress relaxation of near-critical gels} 
  \z{Phys. Rev. E}{64}{021404-1--19}{2001}

\bibitem{BrLo01} 
  \au{K.}{Broderix}, \au{H.}{L\"owe}, \au{P.}{M\"uller}\et\au{A.}{Zippelius}%
    \ti{Critical dynamics of gelation} 
    \z{Phys. Rev. E}{63}{011510-1--17}{2001}

\bibitem{BuHa96}
  \au{A.}{Bunde}\et\au{S.}{Havlin} (Eds.)%
  \bti[2nd rev. and enl. ed.]{Fractals and disordered systems}
  \pub[Chap.~2, 3]{Springer}{Berlin}{1996}

\bibitem{CaLa90}
  \au{R.}{Carmona}\et\au{J.}{Lacroix}%
  \bti{Spectral theory of random Schr\"odinger operators}
  \pub{Birk-h\"auser}{Boston}{1990}

\bibitem{ChCh86}
  \au{J. T.}{Chayes}, \au{L.}{Chayes}, \au{J. R.}{Franz},
  \au{J. P.}{Sethna}\et\au{S. A.}{Trugman}%
  \ti{On the density of states for the quantum percolation problem}
  \z{J. Phys. A}{19}{L1173--L1177}{1986}
    
\bibitem{Chu97}
  \au{Fan R. K.}{Chung}%
  \bti{Spectral graph theory}
  \pub{American Mathematical Society}{Providence, RI}{1997}
    
\bibitem{ChGr00}
  \au{Fan}{Chung}, \au{A.}{Grigor'yan}\et\au{Shing-Tung}{Yau}%
  \ti{Higher eigenvalues and isoperimetric inequalities on Riemannian manifolds
    and graphs}
  \z{Commun. Anal. Geom.}{8}{969--1026}{2000}

\bibitem{Col98}
  \au{Y.}{Colin de Verdi\`ere}%
  \bti{Spectres de graphes}
  \pub*[in French]{Soci\'et\'e Math\'ematique de France}{Paris}{1998}
    
 \bibitem{CvDo95} 
   \au{D.}{Cvetkovi\'c}, \au{M.}{Domb}\et\au{H.}{Sachs}%
   \bti{Spectra of graphs: Theory and applications} 
   \pub{Johann Ambrosius Barth}{Heidelberg}{1995}
  
\bibitem{Gri99}
  \au{G.}{Grimmett}%
  \bti[2nd ed.]{Percolation} 
  \pub{Springer}{Berlin}{1999}

\bibitem{KhSh04}
  \au{O.}{Khorunzhy}, \au{M.}{Shcherbina}\et\au{V.}{Vengerovsky}%
  \ti{Eigenvalue distribution of large weighted random graphs}
  \z{J. Math. Phys.}{45}{1648--1672}{2004}

\bibitem{KiEg72}
  \au{S.}{Kirkpatrick}\et\au{T. P.}{Eggarter}%
  \ti{Localized states of a binary alloy}
  \z{Phys. Rev. B}{6}{3598--3609}{1972}

\bibitem{Kir89}
  \au{W.}{Kirsch}%
  \ti{Random Schr\"odinger operators: a course} 
  In
  \au{H.}{Holden}\et\au{A.}{Jensen} (Eds.)%
  \bti[Lecture Notes in Physics 345.]{Schr\"odinger operators}
  \pub[pp. 264--370]{Springer}{Berlin}{1989}

\bibitem{LeMu03}
  \au{H.}{Leschke}, \au{P.}{M\"uller}\et\au{S.}{Warzel}%
  \ti{A survey of rigorous results on random Schr\"o\-dinger operators for 
    amorphous solids}
  \z{Markov Process. Relat. Fields}{9}{729--760}{2003}.
  A slightly longer and updated version appeared in
  \au{J.-D.}{Deuschel}\et\au{A.}{Greven} (Eds.)%
  \bti{Interacting stochastic systems}
  \pub[pp. 119--151]{Springer}{Berlin}{2005}
 
\bibitem{Moh91}
  \au{B.}{Mohar}%
  \bti{The Laplacian spectrum of graphs, graph theory, combinatorics,
    and applications}
  \pub{Wiley}{NewYork}{1991}

\bibitem{Mul03}
  \au{P.}{M\"uller}%
  \ti{Critical behaviour of the Rouse model for gelling polymers}
  \z{J. Phys. A}{36}{10443--10450}{2003}
  
\bibitem{MuSt05}
  \au{P.}{M\"uller}\et\au{P.}{Stollmann}%
  \ti{Spectral asymptotics of the Laplacian on supercritical bond-percolation
    graphs} 
  E-print math-ph/0506053 (2005)
  
\bibitem{PaFi92}
  \au{L.}{Pastur}\et\au{A.}{Figotin}%
  \bti{Spectra of random and almost-periodic operators}
  \pub{Springer}{Berlin}{1992}

\bibitem{ReSi78}
  \au{M.}{Reed}\et\au{B.}{Simon}%
  \bti{Methods of modern mathematical physics IV: Analysis of operators}
  \pub{Academic}{New York}{1978}
 
\bibitem{ShAh82}
  \au{Y.}{Shapir}, \au{A.}{Aharony}\et\au{A. B.}{Harris}
  \ti{Localization and quantum percolation}
  \z{Phys. Rev. Lett.}{49}{486--489}{1982}

\bibitem{Sim85}
  \au{B.}{Simon}%
  \ti{Lifschitz tails for the Anderson model}
  \z{J. Stat. Phys.}{38}{65--76}{1985}

\bibitem{StAh94}
  \au{D.}{Stauffer}\et\au{A.}{Aharony}%
  \bti[revised 2nd ed.]{Introduction to percolation theory}
  \pub{Taylor and Francis}{London}{1994}  

\bibitem{Sto01}
  \au{P.}{Stollmann}%
  \bti{Caught by disorder: lectures on bound states in random media}
  \pub{Birkh\"auser}{Boston}{2001}

\bibitem{Ves05a}
  \au{I.}{Veseli\'c}%
  \ti{Quantum site percolation on amenable graphs}
  In
  \au{Z.}{Drmac}, \au{M.}{Marusic}\et\au{Z.}{Tutek} (Eds.)%
  \bti{Proceedings of the Conference on Applied Mathematics and Scientific
    Computing}
  \pub{Springer}{Berlin}{2005}
  
\bibitem{Ves05b}
  \au{I.}{Veseli\'c}%
  \ti{Spectral analysis of percolation Hamiltonians}
  \z{Math. Ann.}{331}{841--865}{2005}

\end{thebibliography}
\end{document}